\documentclass[fleqn,usenatbib]{mnras}

\usepackage{newtxtext,newtxmath}

\usepackage[T1]{fontenc}
\usepackage{ae,aecompl}

\usepackage{graphicx}	
\usepackage{amsmath}	
\usepackage{amssymb}	

\title[MTS$-\Gamma$ Correlation]{A MAD Explanation for the Correlation between bulk Lorentz factor and Minimum Variability Timescale}

\author[Authors]{
Nicole Lloyd-Ronning,$^{1,2}$\thanks{E-mail: lloyd-ronning@lanl.gov}
Wei-hua Lei,$^{3}$, Wei Xie$^{4}$
\\
$^{1}$Center for Theoretical Astrophysics, Los Alamos National Lab, Los Alamos, NM USA 87544\\
$^{2}$University of New Mexico-Los Alamos, Los Alamos, NM USA 87544\\
$^{3}$School of Physics, Huazhong University of Science and Technology, Wuhan 430074, China\\
$^{4}$Guizhou Provincial Key Laboratory of Radio Astronomy and Data Processing, Guizhou Normal University, Guiyang, 550001, China. 
}

\date{Accepted XXX. Received YYY; in original form ZZZ}

\pubyear{2017}

\begin{document}
\label{firstpage}
\pagerange{\pageref{firstpage}--\pageref{lastpage}}
\maketitle

\begin{abstract}
 We offer an explanation for the correlation between the minimum variability timescale ($MTS$) in the prompt emission light curve of gamma-ray bursts (GRBs) and the estimated bulk Lorentz factor of these GRBs, in the context of a magnetically arrested disk (MAD) model.  In particular, we show that previously derived limits on the maximum available energy per baryon in a Blandford-Znajek jet leads to a relationship between the characteristic MAD timescale, $t_{MAD}$, in GRBs and the maximum bulk Lorentz factor: $t_{MAD} \propto \Gamma^{-6}$, somewhat steeper than (although within the error bars of) the fitted relationship found in the GRB data.  Similarly, the MAD model also naturally accounts for the observed anti-correlation between $MTS$ and gamma-ray luminosity $L$ in the GRB data, and we estimate the accretion rates of the GRB disk (given these luminosities) in the context of this model.  Both of these correlations ($MTS-\Gamma$ and $MTS-L$) are also observed in the AGN data, and we discuss the implications of our results in the context of both GRB and blazar systems.
\end{abstract}




\section{Introduction}
  Gamma-ray bursts (GRBs) exhibit rapid variability ($\leq 1s$) in their gamma-ray light curves during the first 100 seconds or so. However, the underlying nature of this variability is still an open question. Radially modulated outflow has long been a viable mechanism for reproducing the observed prompt GRB variability  and can reflect the inherent internal variance of the GRB central engine \citep{KPS97, RRF00, NP02, Pesc18}.  However, other models have been invoked to explain the gamma-ray light curve variability, such as angular variation in the jet due to a corrugation instability (e.g. \cite{TG14}), magnetic turbulence in the emitting region \citep{ZY11}, or multiple ``minijets'' producing the prompt gamma-ray emission \citep{BLG16}.  
  
  Under the reasonable assumption that the GRB gamma-ray variability is a direct manifestation of the behavior of the central engine, \cite{lr16} presented a model for the variability in the context of a magnetically arrested disk (MAD; \cite{Nar03}).  In their model, the variability timescale is related to the free-fall time in the disk, in the region over which accretion is stalled or arrested.  For fiducial black hole-disk parameters - particularly the mass of the black hole, the radius to which accretion is arrested and the ``degree'' of arrested-ness - this model reproduces the observed $\sim 1s$ variability timescales of prompt GRB light curves. 
   
    Beyond the variability timescale, however, other properties of the prompt gamma-ray light curve may help uncover the underlying nature of GRB variability.  For example, pulse width distributions \citep{N96, NP02, Q02}, the distribution of intervals between pulses, as well as correlations between timing and other GRB properties (e.g. spectral lag-luminosity correlations \citep{NMB00}, luminosity-variability correlations \citep{FRR00,Reich01}, spectral peak energy-variability correlations \citep{LRRR02}, the relation between the quiescent time before a pulse and the duration of the subsequent pulse \citep{RRM01, NP02, DP07}) may all help shed light on the nature of the variability in gamma-ray bursts.
    
   Recently, \cite{sonbas15} found an anti-correlation between the minimum timescale ($MTS$) of variability in the prompt light curve and the bulk Lorentz factor $\Gamma$ of a GRB. This work was later confirmed by \cite{wu16}. In addition to confirming and refining the correlation seen between $MTS$ and bulk Lorentz factor in GRBs, they extended the work to blazars and found a joint correlation (combining GRB and AGN data) of $MTS \propto \Gamma^{-4.7 \pm 0.3}$.  Both studies also found an anti-correlation between MTS and the isotropic gamma-ray luminosity, $L$, in GRBs. \cite{wu16}, however, extended this study and found the $MTS-L$ correlation also exists in AGN. They found  $MTS \propto L^{-1.0 \pm 0.1}$, for GRB data alone, and $MTS \propto L^{-1.1 \pm 0.1}$ for the GRB and blazar data combined (see their Figure 2).  Although this strongly suggests the same underlying physical processes in blazar and GRB jets, the exact nature of this correlation is, again, not well-established.  
    
    Here, we examine the correlation between bulk Lorentz factor and minimum variability timescale in the context of a magnetically arrested disk model for a GRB central engine. We focus on the $MTS-\Gamma$ correlation but discuss the (presumably related) $MTS-L$ correlation in the context of this model as well.  We apply this model to the GRB data in this paper, but discuss its application to AGN/blazar systems.  Our paper is organized as follows.  In \S~\ref{sec:data}, we describe the data sample of \cite{wu16} and their results for the MTS-$\Gamma$ correlation.  In \S \ref{sec:mad}, we show how this correlation can be reproduced in the context of a MAD model for the GRB central engine, and we present estimates for the accretion rates in GRB disks in the context of this model.  A discussion of the $MTS-L$ correlation, the limitations of our model, and other possible interpretations for this correlation are  discussed in \S ~\ref{sec:disc}. Our conclusions are presented in \S ~\ref{sec:conc}.

\section{Data Sample}
\label{sec:data}
 We use the data sample from \cite{wu16}, who showed that both gamma-ray bursts and blazars exhibit a quantitatively similar correlation between the jet's bulk Lorentz factor $\Gamma$ and the minimum variability timescale (MTS) in their light curves.  The GRB sample is derived originally from that of \cite{sonbas15} who compiled a sample of {\em Swift} and {\em Fermi} gamma-ray bursts for which an $MTS$ and $\Gamma$ could be determined.  \cite{wu16} refined the analysis of the \cite{sonbas15} sample in two ways: 1) Their measurements of $MTS$ are derived from a method using non-decimated Haar wavelets, which is less dependent on the underlying noise level than the \cite{sonbas15} sample (see \cite{wu16, GB14, Gol15} for the details of their methods). 2) Their Lorentz factors are taken from \citet{Lu12} and \citet{Liang15} who use the time of onset of the peak of the afterglow (before the outflow is in the self-similar Blandford-McKee stage) to determine $\Gamma$:
 
 \begin{equation}
 \Gamma  \approx 1.4 \Big[\frac{3E_{\rm \gamma,iso}(1+z)^{3}}{32\pi n m_{\rm p}c^{5} \eta t_{\rm dec}^{3}}\Big]^{1/8}
 \end{equation}
 
\noindent where $t_{\rm dec}$ is the deceleration time, $n$ is the circumburst medium density, $\eta$ is the ratio between gamma-ray and kinetic energy, $m_{\rm p}$ is the proton mass, $c$ is the speed of light, $z$ is the redshift, and $E_{\rm \gamma, iso}$ is the isotropic gamma-ray energy.

   With this updated and improved analysis, they find the correlation for the GRB sample alone is $MTS \propto \Gamma^{-4.8 \pm 1.5}$.  Combining the GRB and blazar data, they find $MTS \propto \Gamma^{-4.7 \pm 0.3}$.

\subsection{Potential Sample Bias}
  It is important to ask - when investigating any correlation - whether there is a potential selection effect artificially producing the purported relation between variables (for a discussion of these issues in gamma-ray bursts, see \citet{LP96, PL96, LPM00, BKB09, D15}).  In the case of the $MTS-\Gamma$ correlation, we need to keep in mind that only those bursts for which an $MTS$ could be determined were selected.  Because a sufficient signal to noise is required, this selects for brighter bursts. In addition, Lorentz factors are obtained only for those bursts which have measurements for the time of onset of the peak of the afterglow, which also potentially selects for brighter bursts.  
  
   Ultimately, then, we need to consider the lower left quadrant in of the MTS$-\Gamma$ plane: lower MTS's are harder to determine and lower $\Gamma's$ correlate with lower luminosities so there may be a selection against observing GRBs in this part of parameter space, which would artificially lead to a stronger correlation than what is observed.  
  
  However, we note that even when populating this part of the $MTS-\Gamma$ plane with data, there is an absence of bursts with high $MTS$ and high $\Gamma$, which is unlikely to be a selection effect and requires explanation.  In other words, some amount of correlation would still exist even if there is a selection effect in the lower left quadrant of the MTS$-\Gamma$ plot. 
  
  In principle, there are non-parametric techniques to estimate the degree of correlation present when selection effects artificially truncate the relevant parameter space (e.g. \cite{LB71,ep98}). In our case, where the truncation is not well-defined and the sample size is relatively small, these techniques have limited application. For the purposes of this paper, we make the assumption that the analysis and results of \citet{sonbas15} and \citet{wu16} are robust within the correlation index error bars of $ \pm 1.5 $, and offer an explanation of these correlations in terms of a magnetically arrested disk.
  

\section{A MAD explanation}
\label{sec:mad}  
  As discussed above, the rapid variability of a prompt gamma-ray burst light curve can reflect the internal variability of the inner engine.  On possibility for this variability is described in \citet{lr16} in the context of a magnetically arrested disk \citep{Nar03}.  In this model, magnetic flux is dragged in by gas as it accretes onto the black hole and this flux is held at the horizon.  If enough flux is built up, it provides magnetic pressure against accreting gas, ''arresting'' the flow within a certain radius $R_{m}$.  The Blandford-Znajek (BZ) process is active during this time, while the magnetic flux is anchored to the black hole, and outflow is launched \citep{bz77,tm12,lei13,lei17}.  Eventually, however, the magnetic field undergoes an interchange instability \citep{ST90, Nar03} or perhaps reconnection events that dissipate the flux and shut off the BZ process until flux can re-accumulate on the black hole.
  
 In this model, the variability timescale is roughly related to the time it take the gas to fall into the black hole, $t_{MAD} \sim R_{m}/(\epsilon_{MAD} v_{ff})$, where $\epsilon_{MAD}$ is a parameter describing the degree of ``arrestedness'' of the flow (essentially consolidating our ignorance of the microphysics into a single parameter) and $v_{ff}$ is the free-fall velocity, $v_{ff} = \sqrt{2GM/R}$.  From equation 5 of \cite{lr16} we see that:  

\begin{equation}
  t_{MAD} \simeq .3s (\frac{10^{-2}}{\epsilon_{MAD}})(\frac{R_{m}}{30R_{g}})^{3/2}(\frac{M}{5M_{\odot}})
\end{equation}
 
 \noindent where $R_{g}$ is the gravitational radius. We can describe the radius to which the flow is arrested $R_{m}$ with the expression from \citet{Nar03} or \citet{lr16}:
 \begin{equation}
 R_{m}/R_{g} \simeq 5(\phi_{29})^{4/3} (M_{5})^{-4/3}(\dot{M}_{-2})^{-2/3}(\epsilon_{MAD,-2})^{2/3} 
 \end{equation}

\noindent where $\phi_{29}$ is the flux on the black hole in units of $10^{29} G cm^{2}$, $M_{5}$ is the mass of the black hole in units of 5 solar masses, $\dot{M}_{-2}$ is the accretion rate in units of $10^{-2} M_{\odot}s^{-1}$ and $\epsilon_{MAD,-2}$ is normalized to $10^{-2}$. Plugging this expression into the equation above for $t_{MAD}$, we find:
\begin{equation}
t_{MAD} \approx 0.02s \phi_{29}^{2} M_{5}^{-1} \dot{M}^{-1}
\end{equation}
\noindent Hence, there exists an inverse relationship between the MAD timescale and accretion rate: $t_{MAD} \propto \dot{M}^{-1}$.  

For the MAD model to be viable, of course, we need a magnetically dominated jet powered by the BZ mechanism.  Recently, \citet{xie17} examined the $MTS-\Gamma$ correlation in the context of both magnetically launched BZ jets and jets driven by neutrino annihilation.  Taking the $MTS$ as the timescale of the viscous instability of a neutrino dominated accretion flow (NDAF, see \citet{pwf99, dpn02, gu06, janiuk07, lei09, xie16, liu17}) disk, and given their expressions for luminosity and baryon loading rate in a BZ jet, they find that the $MTS-\Gamma$ and $MTS-L$ correlations favor a jet driven by the BZ mechanism (see their section 2.3).  

The maximum Lorentz factor of the jet driven by the BZ mechanism is $\Gamma_{max} = \mu_0$, where $\mu_{0}$ is the maximum available energy per baryon in a BZ jet \citep{lei13, lei17, xie17}:
\begin{eqnarray}
\mu_0&\simeq & 1.5\times10^5 A^{-23/30}B^{33/40}C^{-7/120} f_\text{p,-1}^{1/2}\nonumber\\
&\times& \theta_\text{j,-1}^{-1} \theta_\text{B,-2} \alpha_{-1}^{-23/60}\epsilon_{-1}^{-5/6}r_{z,11}^{-1/2} a^2 X(a) \nonumber\\
&\times& \left(\frac{R_\mathrm{ms}}{2}\right)^{-1/120}\left(\frac{\xi}{2}\right)^{-1/120}\left(\frac{m}{3}\right)^{11/20} \dot{m}_{-1}^{1/6},
\label{eq:mu0}
\end{eqnarray}
\noindent where $f_\mathrm{p}$ denotes the fraction of the protons in the wind, $r_z$ is the distance from the BH in the jet direction, $\theta_\mathrm{B}$ is introduced here to reflect the fact that only the protons with small ejected angle ($\le \theta_\mathrm{B}$) with respect to the field lines can come into the disk atmosphere. The parameter $\xi \equiv r/r_{\rm ms}$ is the disk radius in terms of $r_{\rm ms}$, $\epsilon\simeq(1-E_\mathrm{ms})$ denotes the neutrino emission efficiency, $E_\mathrm{ms}=(4\sqrt{R_\mathrm{ms}}-3a)/\sqrt{3}R_\mathrm{ms}$ is the specific energy at the ISCO. The parameters A, B, C and D are the relativistic correction factors for a disk around a Kerr BH \citep{RH95}. The parameter $a$ is the spin of the black hole, and $X(a)$ is defined as $X(a) = F(a)/(1+\sqrt{1-a^2})^{2}, F(a) = [(1+q^2)/q^2][(q+1/q)\rm{arctan}(q -1)]$, and $q = a/(1+\sqrt{1-a^2})$. 

Our mass loading model (leading to the maximum Lorentz factor $\mu_{o}$ above) is formulated in the context of a Blandford-Znajek jet supported by a disk dominated by neutrino cooling processes. The black hole magnetosphere has a limiting charge density defined by the force-free condition of a BZ jet, which leads to a minimum baryon loading rate (e.g. equation 27 of \citet{lei13}). The dominant source of mass loading in the jet comes neutron drift from a neutrino-driven wind in the disk (the magnetic field will prevent protons from entering the jet). These neutrons are then converted to protons via positron capture and proton$-$neutron inelastic collisions in the jet.  These conditions are expected to be particularly relevant for hyper$-$accreting GRB disks, and may also hold for at least some AGN/blazar disk systems. We refer the reader to \S 3 of \cite{lei13} for further details of our baryon loading model.

Applying this proportionality to our MAD timescale, and taking $\Gamma \approx \mu_{0}$, we find:
\begin{equation}
t_{MAD} \propto \Gamma^{-6}
\end{equation}
 
 Using $t_{MAD}$ as our variability timescale, this equation is consistent (within the error bars) with the quantitative proportionality found in the \cite{wu16} sample between minimum variability timescale and Lorentz factor in the GRB data, $MTS \propto \Gamma^{-4.8 \pm 1.5}$. Note that in our substitution, we set $\Gamma$ equal to the maximum Lorentz factor $\mu_{o}$ from \cite{lei13}.  In reality, the Lorentz factor may not reach this maximum value and and will lead to a softening and/or scatter in the correlation, in line with what is observed. The additional parameters present in equation 5 will also lead to scatter and a softening of the actual observed correlation between MTS and $\Gamma$.  Finally, our MAD timescale is roughly the largest timescale in which we expect variability in the disk in a MAD state.  Lower values of $t_{MAD}$ can also serve to soften the actual (observed) correlation.
 
 \subsection{Accretion Rates in the MAD Model}
 Under the assumption that the GRB luminosity is powered by a BZ jet, we can estimate the necessary accretion rate to reproduce the observed GRB luminosity (some fraction $\eta$ of the BZ luminosity).
Combining equations 3 and 4 of \citet{lr16} for the GRB luminosity in MAD scenario, we have
 \begin{equation}
 L_{GRB} = \eta L_{BZ} = \eta \frac{kfc^{2}}{32\pi}a^{2}\epsilon_{MAD}^{-1}\dot{M}
 \end{equation}
 where, again, $\epsilon_{MAD}$ is the degree of "arrestedness" of the accretion flow, $a$ is the spin parameter of the black hole, $k$ is a geometrical factor $\sim 0.05$ and $f$ is of order unity.  Once again, our ignorance of the microphysics (which may in principle play a role in the jet launch and resultant luminosity) is parameterized by our efficiency factors $\eta$ and $\epsilon_{MAD}$.  These factors could in principle depend on accretion rate themselves.  However, for our purposes, we utilize the standard assumption that $\eta$ and $\epsilon_{MAD}$ are constants (independent of accretion rate).
 Then, solving for $\dot{M}$, we have
 \begin{equation}
 \dot{M} = (0.1 M_{\odot} s^{-1}) a^{-2}\eta^{-1}\epsilon_{MAD} (L_{GRB}/10^{52}erg s^{-1})
 \end{equation}
 
 Table~\ref{tab:data} gives the accretion rates for the GRBs in the \citet{wu16} sample for an assumed $a \sim 1$, $\eta=0.1$, $\epsilon_{MAD}=0.001$ (see \citet{lr16} for a explanation of the choice of these parameters).
 
\begin{figure*}
\centering
\includegraphics[width=5.5in]{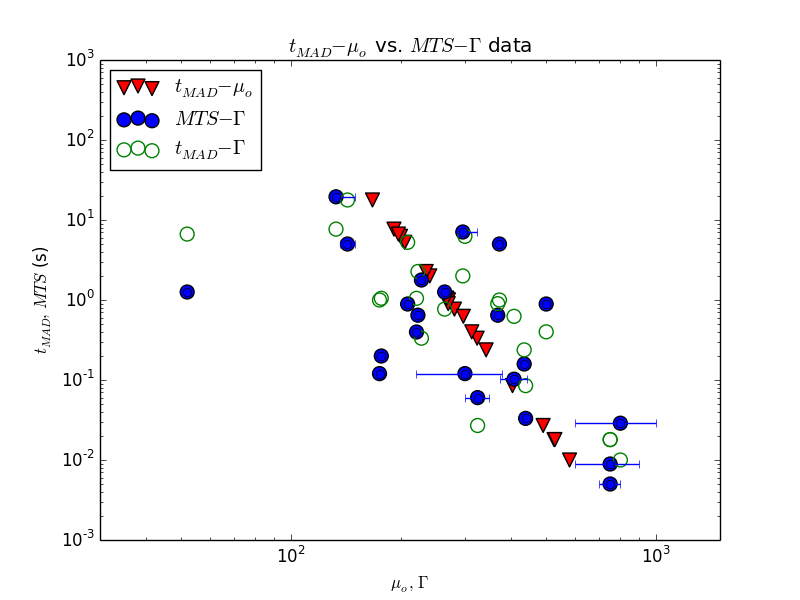}
\caption{Normalized MAD timescale $t_{MAD}$ vs. predicted maximum Lorentz factor $\mu_{o}$ (red triangles), $t_{MAD}$ vs. observed/fitted Lorentz factor $\Gamma$ (open green circles), and observed minimum variability timescale as a function of fitted/observed Lorentz factor $\Gamma$ (blue circles).}
\label{fig:tminGamma}
\end{figure*}

 In Figure~\ref{fig:tminGamma}, we plot a normalized MAD timescale from equation 4 vs. the predicted maximum Lorentz factor $\mu_{o}$ from equation 5 (red triangles), varying only the accretion rate (derived from the equation 9 and using the observed GRB luminosities). Superposed on this (green open circles) is the MAD variability timescale $t_{MAD}$ (again employing accretion rates inferred from observed GRB luminosities in a Blandford-Znajek model) vs. the observed/fitted Lorentz factor $\Gamma$ of the \citet{wu16} sample.  For comparison, we also plot the \citet{wu16} data (blue circles) for the GRB sample (see their Table 1 and also our Table~\ref{tab:data} below). It is evident our MAD model reproduces the observed $MTS-\Gamma$ correlation.
 
\section{Discussion}
\label{sec:disc}
We have presented an explanation for the $MTS-\Gamma$ relation seen in both GRBs and AGNs, and applied it to the existing GRB data.  We note that the $MTS$-Luminosity relation (also presented in \cite{sonbas15} and \cite{wu16}) is also a natural consequence of the MAD model.  In the MAD model, $L \propto \dot{M}$.  Because our MAD variability timescale $t_{MAD}$ is proportional to $\dot{M}^{-1}$, this relation is naturally explained in the context of the MAD model: $MTS \approx t_{MAD} \propto L^{-1}$ (see equation 4).  

Our model makes several simplifying assumptions. We again point out that we are working under the premise that the observed GRB variability reflects the central engine variability. We assume enough magnetic flux can be built up/is present at the horizon to arrest the accretion, such that a MAD disk is viable.  We then assume an interchange instability or reconnection events will allow the field to diffuse/dissipate and this occurs on the timescale of the free fall time in the magnetically arrested disc, in line with \citet{ST90} who showed the growth rate of the interchange instability occurs on the order of the free fall time if the ratio of surface density to poloidal field changes significantly enough as a function of radius. Although our MAD picture circumvents the details of the microphysics of these processes (with a generic parameterization $\epsilon_{MAD}$), it captures the global behavior of the disk under the assumptions mentioned above.  Finally, we have focused on the accretion rate as the determining variable in examining these correlations.  As evident in equation 5, other parameters come into play and will add scatter to the correlation.  

 We note that although this model can be extended to AGN, there have been a number of blazars with observed variability timescales that are comparable to, or shorter than, the light crossing times of their central black holes (for example, see Figure 1 of \cite{2015vb}).  These observed MTSs are shorter than our MAD timescale, $t_{MAD}$, and therefore cannot be explained in this context.  Variability on these extremely short timescales is instead likely due to small scale inhomogeneities in the jet (i.e. local emission sites), potentially far from the central black hole \citep{BFR08,2015vb}.

Other studies have examined prompt GRB light curve variability in other contexts. \citet{xie17} attribute the variability to a viscous instability in a neutrino dominated accretion flow (NDAF). \citet{Cao14} also consider an NDAF in a MAD disk, but attribute the minimum variability timescale to the magnetic field diffusion time (their equation 25).  An alternative explanation (see, e.g. \citet{wu16} ) for the $MTS-\Gamma$ anti-correlation is due simply to pair opacity effects - a highly variable light curve is only possible if the optical depth to pair production is low enough. The variability sets the distance scale over which the photons are emitted - if this distance is too small, the region will be optically thick to pair production.  This optical depth is reduced for  relativistic outflows ($\tau_{\gamma \gamma} \propto \Gamma^{-(2\beta +2)}$, where $\beta$ is the high energy spectral index of the photon spectrum, typically $\approx 2$); hence, higher $\Gamma$ outflows would naturally allow for smaller variability timescales (note that this transparency condition gives only a lower limit to the Lorentz factors).
 
 Finally, another popular scenario for the jet production is neutrino annihilation \citep{pwf99, dpn02, gu06, janiuk07, lei09, xie16, liu17}, in which the maximum Lorentz factor of the jet is determined by the dimensionless ``entropy'' parameter, $\eta_0$, i.e. \citep{lei13, lei17, xie17},
\begin{eqnarray}
\eta_0& =  & 50 A^{-1.13} B^{1.35} C^{-0.22} \theta_{\rm j,-1}^{-2} \alpha_{-1}^{-0.57} \epsilon_{-1}^{-1.7} \left(\frac{\xi}{2}\right)^{-0.32} \nonumber \\
& \times & \left(\frac{ R_{ms}}{2}\right)^{-5.12} \left(\frac{m}{3}\right)^{-0.6} \dot{m}_{-1}^{0.58}
\label{eq:eta}.
\end{eqnarray}
As shown in \citet{xie17}, this model can hardly reproduce the observed $MTS-L$ correlation. It is worth mentioning that they defined $MTS$ as the viscous timescale due to the viscous instability. If, on the other hand, the GRB variability is the result of the MAD timescale, this model will then predict $t_{MAD} \propto \Gamma^{-1.7}$, which is still inconsistent with the observations. Therefore, our MAD explanation also favors the scenario in which the jet is driven by BZ mechanism.


\section{Conclusions}

 We have shown that the variability timescale in a MAD model as defined in \citet{lr16} can reproduce the observed minimum variability timescale ($MTS$) - bulk Lorentz factor ($\Gamma$) correlation as well as the $MTS$ - luminosity ($L$) relation observed in the long GRB data, and can also be applied/extended to AGN data. In particular, we use the relationship between Lorentz factor and accretion rate defined in \cite{lei13} to show that $MTS \sim t_{MAD} \propto \Gamma^{-6}$ and $MTS \sim t_{MAD} \propto L^{-1}$.  Of course variation in other physical parameters will serve to add scatter to the correlation, but the general relation is present in this model.
 
 This paradigm can apply to any black hole-accretion disk system in which a magnetically arrested disk model might apply (including potentially short GRBs as well).  It has long been noted that GRB and AGN systems are similar - scaled by the mass of the central black hole - with similar physics for the disk physics, the launch of the jet, and dissipation mechanisms within the jet and MAD models can be applied to both GRB and AGN systems (see, e.g., \citet{MIR04,BK08,Zam14,BT16}). The quantitative similarity of the $MTS-\Gamma$ and $MTS-L$ correlations in both AGN and GRBs support this point and a MAD model is a natural explanation for the existence of these correlations in both types of systems.  Future observations will more securely establish the existence of these correlations, putting them on firmer footing, mitigating any potential selection effects, and allowing us to further test the MAD model as the underlying source of these correlations.

\label{sec:conc}



\begin{table}
	\centering
	\caption{Data sample with $MTS$ values, luminosities, and estimated accretion rates in our MAD model.}
	\label{tab:data}
	\begin{tabular}{lccccr}  
		\hline
		GRB & log(MTS/s)  & $\Gamma$ & log($L_{GRB}/erg \ s^{-1}$)  & $\dot{M} (M_{\odot}\ s^{-1})$\\
		\hline
	 060210 &     0.10 &    264 &     51.97 &    0.13    \\
      060607A &    0.85 &    296 &      51.57 &   0.05  \\
      061007 &    -0.80 &      436 &      52.50 &    0.42   \\
      061121 &    -0.92 &    175 &      51.88 &     0.10   \\
      070318 &     0.70 &      143 &      50.62 &   0.0056   \\
      071010B &   -0.05 &      209 &      51.15 &    0.019   \\
     071031 &      1.29 &      133 &     50.98 &   0.013   \\
      080319B &     -1.40 &      -  &    52.65 &    0.60  \\
      080319C &     0.25 &      228 &      52.35 &     0.30   \\
      080804 &    -0.15 &      -  &    52.43 &     0.36  \\
      080810 &    -0.99 &      409 &      52.08 &     0.16   \\
      090102 &     -1.48 &     440 &     52.94 &      1.18  \\
      090323 &    -0.99 &      -    &  53.08 &      1.63   \\
      090424 &    -0.92 &      300 &      51.08 &    0.016   \\
      090510 &     -2.30 &      750 &      53.61 &      5.51   \\
      090618 &   -0.55 &        -     &   52.31 &    0.27   \\
      090812 &   -0.05 &      501 &     52.27 &     0.25   \\
      090902B &     -2.05 &      750 &      53.62 &      5.64  \\
      090926A &   -1.54 &      800 &      53.87 &     10.035   \\
      091003 &     -1.38 &      -   &   51.93 &     0.11   \\
      091029 &    -0.40 &      221 &      51.85 &   0.095   \\
      100414 &     -1.59 &     -   &  52.75 &    0.76   \\
      100621A &     0.10 &      52 &      51.04 &    0.015   \\
      100728B &     0.70 &      373 &      51.89 &     0.10   \\
      100906A &    -0.19 &      369 &      51.90 &     0.11   \\
      110205A &    -0.70 &      177 &      51.85 &   0.095   \\
      110213A &    -0.19 &      223 &      51.52 &    0.044  \\
      130427A &     -1.22 &      325 &      53.43 &      3.64   \\
		\hline
	\end{tabular}
\end{table}

\section*{Acknowledgements}
We are very grateful to the referee for comments and suggestions that improved this manuscript.  We also thank Josh Dolence for discussions about MAD disks.  Work at LANL was done under the auspices of the National Nuclear Security Administration of the US Department of Energy at Los Alamos National Laboratory LA-UR--18-21316. WHL acknowledges support by the National Basic Research Program ('973' Program) of China (grants 2014CB845800), the National Natural Science Foundation of China under grant 11773010.



\bibliographystyle{mnras}
\bibliography{refs} 








\bsp	
\label{lastpage}
\end{document}